\documentclass{article}
\usepackage{spconf,amsmath,graphicx}
\usepackage{url}
\usepackage{psfrag}
\usepackage{pstool}
\usepackage{multirow}
\usepackage{cite}

\title{Modeling the Energy Consumption of the HEVC Software Encoding Process using Processor events}
\name{{ Geetha Ramasubbu \ Andr\'e Kaup \ Christian Herglotz }}   \address{Multimedia Communications and Signal Processing, \\ Friedrich-Alexander University Erlangen-Nürnberg (FAU), Erlangen, Germany\\ \{geetha.ramasubbu, andre.kaup, christian.herglotz\}@fau.de. \thanks{This work was funded by the Deutsche Forschungsgemeinschaft (DFG, German Research Foundation) – Project-ID 447638564.}}
\usepackage{textcomp}
\usepackage[absolute]{textpos}
\def\BibTeX{{\rm B\kern-.05em{\sc i\kern-.025em b}\kern-.08em
    T\kern-.1667em\lower.7ex\hbox{E}\kern-.125emX}}
\newcommand{\copyrightstatement}{
    \begin{textblock}{15}(0.3,0.2)    
         \noindent
         \centering
         \textblockcolour{white}
         \footnotesize
         \copyright 2024 IEEE. Personal use of this material is permitted. Permission from IEEE must be obtained for all other uses, in any current or future media, including reprinting/republishing this material for advertising or promotional purposes, creating new collective works, for resale or redistribution to servers or lists, or reuse of any copyrighted component of this work in other works.
    \end{textblock}
}
\begin{document}

\copyrightstatement
%
\maketitle
\vspace{-0.35cm}
\begin{abstract}
\vspace{-0.25cm}
Developing energy-efficient video encoding algorithms is highly important due to the high processing complexities and, consequently, the high energy demand of the encoding process. To accomplish this, the energy consumption of the video encoders must be studied, which is only possible with a complex and dedicated energy measurement setup. This emphasizes the need for simple energy estimation models, which estimate the energy required for the encoding. Our paper investigates the possibility of estimating the energy demand of a HEVC software CPU-encoding process using processor events. First, we perform energy measurements and obtain processor events using dedicated profiling software. Then, by using the measured energy demand of the encoding process and profiling data, we build an encoding energy estimation model that uses the processor events of the \textit{ultrafast} encoding preset to obtain the energy estimate for complex encoding presets with a mean absolute percentage error of 5.36\% when averaged over all the presets. Additionally, we present an energy model that offers the possibility of obtaining energy distribution among various encoding sub-processes.
\end{abstract}
\begin{keywords}
video coding, energy efficiency, HEVC, x265 presets, and energy estimation.
\end{keywords}
\vspace{-0.25cm}
\section{Introduction}
\vspace{-0.25cm}
\label{sec:intro}
The convenience of the Internet and mobile devices has increased Internet video traffic lately \cite{ciscoWhitepaper1823}. In addition, video-focused social networking services are growing, accounting for further video traffic increases, and \cite{ciscoWhitepaper1823} emphasizes the enormous storage costs, space needs, and increased server-side energy consumption for video content creation. Furthermore, the compression methods used for encoding have evolved immensely in recent years. Modern codecs provide many compression methods, increasing the encoders' processing complexity \cite{VVCComplexity}, resulting in a considerable increase in the energy demand. 

For several reasons, research on the energy consumption of video encoders is helpful. Firstly, portable devices are limited in terms of battery capacity. The energy requirements for video encoding are significant, which is a problem for battery-powered devices, where the battery drains fast due to increased energy requirements \cite{Sharrab13}. Secondly, the total energy consumption of current encoding and decoding systems is globally significant, as online video contributes to $1\%$ of global $\mathrm{CO}_2$ emissions \cite{shiftProject19}. According to \cite{shiftProject19}, there is a considerable energy demand for using various digital equipment in the video processing pipeline and producing such devices \cite{shiftProject19}. Thirdly, most video-on-demand services use massive server farms in data centers for encoding \cite{Herglotz2022_SweetStreams}. A recent study states that such data centers, which predominantly perform CPU encoding, currently consume about 3\% of global power consumption, which is expected to reach more than 1,000 TW by 2025, equivalent to 1 trillion tons of coal \cite{HuaweiReport}. To enable energy-aware video-based services in modern video communication applications, we need robust and energy-efficient video codecs that optimize energy. 

Few works have explicitly addressed the encoder's processing energy, such as \cite{Rodriguez15}, which establishes a relationship between the quantization, spatial information, and coding energy for the intra-only High Efficiency Video Coding (HEVC) encoder. However, it did not consider the presets that quantify the encoding speed and compression performance. A study on detailed energy consumption of the x265 encoder is presented in \cite{encEnergy2017}. In \cite{katsenou22}, energy-rate-quality trade-offs of state-of-the-art video codecs were studied. Furthermore, Mercat et al. measured the energy of a software encoder for different sequences and encoding configurations and presented various energy reduction opportunities\cite{Mercat17}. In a carbon footprint-limited future, the energy efficiency of video coding is more relevant. The complex and laborious nature of energy measurements is a limitation in searching for energy-efficient algorithms. Therefore, we need simple energy estimation models to overcome the drawback of time-consuming measurements. There is extensive literature on estimating the energy demand of the decoding process, such as \cite{Li12, Raoufi13, Herglotz18c}. However, only some recent works explicitly addressed the estimation of the energy demand of the encoding process. For instance, \cite{Rodriguez15} performs encoding energy estimation using quantization parameter (QP), albeit restricted to all-intra encoding.

Furthermore, \cite{Ramasubbu22} introduces a model that uses the encoder processing time to estimate the energy consumption of the x265 encoder. However, the energy encoding process has to be performed on a target device for estimation. Moreover, \cite{Ramasubbu22} also introduces a practical and lightweight encoding time-based encoding energy estimator, which uses the \textit{ultrafast} (UF) preset's encoding time to enable prior estimation of the encoding energy demand of the other presets. However, the estimation error is more than 15\%, especially for complex presets. Lastly, \cite{Ramasubbu22_2} introduces a bit stream feature-based energy estimator that uses the compressed video bit stream features (BSF) obtained using a bit stream analyzer \cite{Ramasubbu22_2} after encoding to estimate the energy using bit streams. Even though this model makes accurate estimations, it is limited to specific applications, such as estimating the energy of crowd-sourced video data, since the bit stream is required for estimation. To this end, this work explores the feasibility of accurately estimating the HEVC software encoders’ energy demand using processor events (PEs).

In order to achieve this, we perform energy measurements and obtain processor events (PEs) using a dedicated open-source profiling software \cite{valgrind} and then study the relationship between PEs and encoding energy, followed by a study of feasibility for estimating the HEVC software encoder's energy demand. Ultimately, we propose two simple models to efficiently estimate encoding energy without measuring it directly before and after the encoding process using certain PEs obtained using existing open-source profiling tools. The models proposed in this work have two practical applications: to estimate the energy for CPU encoding, for example, in data centers, and study the energy distribution of the encoding process, i.e., to obtain the energy of encoding sub-processes.

The rest of this paper is structured as follows: Section \ref{sec:setup} presents the experimental setup used to measure the energy demand of the encoding process, along with sequences used, as well as encoding configurations, then explains the profiling, and further examines the relation between PEs and energy consumed during the encoding process. Further, Section \ref{sec:proposal} introduces the proposed encoding energy estimation models. Section \ref{sec:eval} introduces the evaluation method, discusses the results, and presents an examplary application of the proposed models. Lastly, Section \ref{sec:concl} concludes this work.
\vspace{-0.25cm}
\section{Experimental Setup and Analysis}
\vspace{-0.25cm}
\label{sec:setup}
Our work uses the x265 encoder implementation \cite{x265} on an Intel Xeon processor to perform multi-core encoding. We encode the first eight frames of each sequence at various x265 presets $\textit{ultrafast}$, $\textit{superfast}$, $\textit{veryfast}$, $\textit{faster}$, $\textit{fast}$, $\textit{medium}$, $\textit{slow}$, $\textit{slower}$, $\textit{veryslow}$ and various Constant Rate Factor (CRF) values, 18, 23, 28, 33. In contrast to the previous works \cite{Ramasubbu22}, and \cite{Ramasubbu22_2}, the x265 encoder used in this work is built using Netwide Assembler (NASM), i.e., it uses Single Instruction Multiple Data (SIMD) instructions. 

We consider 22 sequences from the JVET common test conditions \cite{CTC} with various sequence characteristics such as frame rate, resolution, and content. Ultimately, we generated 792-bit streams and further used them to evaluate the models. In addition, we record the QP, encoding time, and bit stream features to enable a comparison with the estimation models from the literature \cite{Rodriguez15}, \cite{Ramasubbu22}, and \cite{Ramasubbu22_2}. 
As in \cite{Herglotz18c}, we describe the energy demand of the encoding process using two consecutive measurements. First, the total energy consumed during the encoding process is described as follows: 
\begin{equation} \label{totalenergy} \small E_{\mathrm {total}} = \int _{t=0}^{T} P_{\mathrm {total}}(t)dt, \vspace{-0.15cm}\end{equation}
where $T$ is the duration of the encoding process and $P_{\mathrm {total}}$ is the total power consumed while encoding. Then, the energy consumed in idle mode over the same encoding duration $T$ is described as follows:
\begin{equation} \label{idle} \small E_{\mathrm {idle}} = \int _{t=0}^{T} P_{\mathrm {idle}}(t)dt, \vspace{-0.15cm} \end{equation}
where $P_{\mathrm {idle}}$ is the power consumed by the device in idle mode. Ultimately, the encoding energy $E_{\mathrm {enc}}$ is the difference between these two measurements.
In this work, we used a desktop PC with an Intel Xeon CPU with 16 cores and CentOS 8 as an operating system (OS). We employed the integrated power meter in the Intel CPUs, Running Average Power Limit (RAPL) \cite{RAPLinAction2018}, that directly returns aggregated energy values $E_{\mathrm {total}}$ and $E_{\mathrm {idle}}$. Additionally, we perform the confidence interval test proposed in \cite{Bendat1971}, to ensure the statistical significance of the measured encoding energies as in \cite{Ramasubbu22}.

\begin{table}[]
\small
\centering
\caption{Recorded PEs by Valgrind with the Pearson Correlation Coefficient (PCC), a measure of the linear relation between recorded PEs and encoding energy.}
\begin{tabular}{|l|l|l|}
\hline
\textbf{ID} & \textbf{PEs} & \textbf{PCC} \\ \hline
1 & I cache reads, Ir & 0.996 \\ \hline
2 & D cache reads, Dr & 0.995 \\ \hline
3 & D cache writes, Dw & 0.994 \\ \hline
4 & I1 cache read misses, I1mr & 0.985 \\ \hline
5 & D1 cache read misses, D1mr & 0.977 \\ \hline
6 & D1 cache write misses, D1mw & 0.970 \\ \hline
7 & LL cache instruction read misses, ILmr & 0.710 \\ \hline
8 & LL cache data read misses, DLmr & 0.664 \\ \hline
9 & LL cache data write misses, DLmw & 0.702 \\ \hline
10 & Conditional branches executed, Bc & 0.994 \\ \hline
11 & Conditional branches mispredicted, Bcm & 0.982 \\ \hline
12 & Indirect branches executed, Bi & 0.986 \\ \hline
13 & Indirect branches mispredicted, Bim & 0.980 \\ \hline
\end{tabular}
\label{tab:PETable1}
\vspace{-0.25cm}
\end{table}

Furthermore, we employed dedicated open-source profiling software, Valgrind, to obtain the PEs \cite{valgrind}. Valgrind analyzes any desired process for instructions, memory accesses, memory leaks, cache misses, and other processor events. In this work, we take the HEVC encoding process and profile it with the Cachegrind tool of the Valgrind framework \cite{callgrindPaper}. The Cachegrind simulates encoding processes' interaction with the machine's cache hierarchy, independent first-level instruction, and data caches (I1 and D1) backed by a second-level cache (L2) \cite{callgrindPaper}. 
Yet, some modern machines have three or four levels of cache, and Cachegrind can auto-detect the cache configuration for these machines. Hence, Cachegrind simulates the first-level (I1 and D1) and last-level (LL) caches. This is because the last-level cache influences runtime, as it masks access to the main memory. Also, the L1 caches often have low associativity, so simulating them can detect processes' bad interaction with the cache. Even though the application of this profiling tool is straightforward, it slows the original process by a factor of 5, depending on the encoding preset. Therefore, for energy measurements, we used the encoding process without profiling. All the recorded PEs are tabulated in Table \ref{tab:PETable1}. 

To illustrate the relationship between the PEs and the encoding energy, we calculate the Pearson Correlation Coefficient (PCC) proposed in \cite{Bendat1971} to express the correlation in numbers. The correlation between the encoding energy and all considered PEs (PCC) are listed in Table \ref{tab:PETable1}. We can observe that the encoding energy correlates to most of the PEs except the last-level cache misses, which have a low correlation. In addition, from Table \ref{tab:PETable1}, we can observe that the highest correlations are obtained for the number of instructions (Ir) followed by the data reads (Dr), date writes (Dw), Conditional branches executed (Bc), Indirect branches executed (Bi), Conditional branches mispredicted (Bcm), and Indirect branches mispredicted (Bim).
\section{Proposed Encoding Energy Estimation Models}
\label{sec:proposal}
Based on the above observations, we propose a PE-based posterior estimation model, i.e., we estimate the encoder processing energy after the encoding process. The model exploits the linear relationship between encoding energy and PEs, such that the energy can be estimated by
\begin{equation} \label{complexity} \small\hat E_{\mathrm {enc}}=\sum_{\forall \text{\textit{i}}}n_{\textit{i}}\cdot e_{\text{\textit{i}}},\vspace{-0.15cm} \end{equation}
where ${\text{\textit{\textit{i}}}}$ denotes the processor event ID as mentioned in Table \ref{tab:PETable1}, and $n_{\text{i}}$ is the number of occurrences of the respective processor event (PE). The parameter $e_{\text{\textit{i}}}$ can be interpreted as a constant mean energy required to execute a certain PE. It must be noted that the estimates can only be obtained when the encoding process is executed once, as the encoding process needs to be profiled by Valgrind. Therefore, we adjust model \eqref{complexity} to allow prior estimation, i.e., energy estimation, without executing the encoder. In this context, we interpret encoding at the \textit{ultrafast} preset as a preprocessing step for slower presets and investigate the modeling of the \textit{utrafast} presets' PEs on encoding energies of all the presets, as the \textit{ultrafast} PEs are relatively less costly to obtain when compared to that of other presets. When we replace the PEs of respective presets with the PEs of \textit{ultrafast} preset, we can also observe similar linear behavior. Therefore, we adapt the model in \eqref{complexity} to enable prior estimation of energy consumption of the encoding process using the PEs of \textit{ultrafast} preset as the UF PE-based prior estimation model as follows:
\begin{equation} \label{complexityUF}\small\hat E_{\mathrm {enc}}=\sum_{\forall \text{\textit{i}}}n_{\text{\textit{i},UF}}\cdot e_{\text{\textit{i}}},\vspace{-0.15cm} \end{equation}
where ${\text{\textit{i}}}$ denotes ID as mentioned in Table \ref{tab:PETable1}, and $n_{\text{\textit{i},UF}}$ is the number of occurrences of the respective \textit{ultrafast} PE. $e_{\text{\textit{i}}}$ is the model parameters later obtained by training as explained in Section \ref{sec:eval}. 

Depending on the operating system, CPU architecture, and encoding algorithms, the energy consumed during the encoding process and the PEs may vary. However, we have a strong indication that the linear relationship between the energy demand of the encoding process and the PEs remains the same. Therefore, the proposed models can be extended to different operating systems, CPU architecture, and encoding algorithms with differing model parameters. This claim is supported by related work on decoding energy \cite{Herglotz17a} and \cite{Kraenzler2023}, which showed that this kind of model is valid for different operating systems, CPU architectures, and codecs.
\section{Evaluation}
\label{sec:eval}
We evaluate the proposed models using the mean absolute percentage error (MAPE), as we strive to estimate the encoding energy accurately independent of the absolute energy, which varies by several orders of magnitude. Thus, we calculate the absolute percentage error of the measured encoding energy concerning the estimated encoding energy for a single bit stream $b$, i.e., each bit stream $b$ corresponds to a single input sequence coded at a specific CRF, and preset $X$. Then, we calculate the mean absolute percentage error ${{\epsilon}_{X}}$ for each preset $X$ over $B$ bit streams to obtain the overall estimation error as MAPE for each preset as follows:
\begin{equation}
\small
    {\epsilon}_{X}= \frac{1}{B}\sum_{b=1}^{B} \vert r_{X,b} \vert, 
    \vspace{-0.15cm}
\end{equation}
where $r_{X,b}$ is the percentage errors for a given preset $X$ and bit stream $b$ and is given as:
\begin{equation}
\small
    r_{X,b}=\left(\frac{\hat E_{\mathrm {enc}}- E_{\mathrm {enc}}}{E_{\mathrm {enc}}}\right) \cdot 100, 
\vspace{-0.15cm}
\end{equation}
where $\hat E_{\mathrm {enc}}$ is the estimated encoding energy and $ E_{\mathrm{enc}}$ the measured encoding energy. To determine the model parameters $e_{\textit{i}}$ for each preset, we perform a least-squares fit using a trust-region-reflective algorithm as presented in \cite{Coleman96}. We use the measured energies for a subset of the sequences referred to as the training set and their corresponding variables as input, which are the PEs. As a result,  we obtain the least-squares optimal parameters for the input training set, where we train the parameters such that the mean relative error is minimized. Ultimately, these model parameters validate the model's accuracy on the remaining validation sequences. The training and validation data set are determined using a ten-fold cross-validation proposed in \cite{Zaki14}. 
\begin{table}[]
\small
\centering
\caption{MAPE for all the posterior estimation models and the confidence interval of percentage errors of the proposed model. The minimum MAPE across all models are in bold.}
\label{tab:posterior}
\begin{tabular}{|l|rrrl|}
\hline
\multirow{3}{*}{\textbf{\begin{tabular}[c]{@{}l@{}}Encoder \\ Preset\end{tabular}}} &
  \multicolumn{4}{c|}{\textbf{Posterior Estimation models}} \\ \cline{2-5} 
 &
  \multicolumn{1}{c|}{\textbf{\begin{tabular}[c]{@{}c@{}}Time\\ \cite{Ramasubbu22}\end{tabular}}} &
  \multicolumn{1}{c|}{\textbf{\begin{tabular}[c]{@{}c@{}}BSF\\ \cite{Ramasubbu22_2}\end{tabular}}} &
  \multicolumn{2}{c|}{\textbf{\begin{tabular}[c]{@{}c@{}}PE\\ \eqref{complexity}\end{tabular}}} \\ \cline{2-5} 
 &
  \multicolumn{3}{c|}{\textbf{MAPE, }$\boldmath{{\epsilon}_{X}}$ \textbf{(\%)}} &
  \multicolumn{1}{c|}{\boldmath{$CI_{X}$\textbf{(\%)}} } \\ \hline
\textit{ultrafast} &
  \multicolumn{1}{r|}{5.96} &
  \multicolumn{1}{r|}{5.05} &
  \multicolumn{1}{r|}{\textbf{4.38}} &
  {[}-1.51, 0.84{]} \\ \hline
\textit{superfast} &
  \multicolumn{1}{r|}{\textbf{4.23}} &
  \multicolumn{1}{r|}{5.03} &
  \multicolumn{1}{r|}{4.65} &
  {[}-1.57, 1.11{]} \\ \hline
\textit{veryfast} &
  \multicolumn{1}{r|}{5.38} &
  \multicolumn{1}{r|}{6.06} &
  \multicolumn{1}{r|}{\textbf{4.40}} &
  {[}-1.28, 1.16{]} \\ \hline
\textit{faster} &
  \multicolumn{1}{r|}{5.19} &
  \multicolumn{1}{r|}{6.37} &
  \multicolumn{1}{r|}{\textbf{4.47}} &
  {[}-1.25, 1.22{]} \\ \hline
\textit{fast} &
  \multicolumn{1}{r|}{5.16} &
  \multicolumn{1}{r|}{6.54} &
  \multicolumn{1}{r|}{\textbf{4.54}} &
  {[}-1.25, 1.21{]} \\ \hline
\textit{medium} &
  \multicolumn{1}{r|}{\textbf{4.62}} &
  \multicolumn{1}{r|}{7.19} &
  \multicolumn{1}{r|}{4.66} &
  {[}-1.46, 1.29{]} \\ \hline
\textit{slow} &
  \multicolumn{1}{r|}{\textbf{3.97}} &
  \multicolumn{1}{r|}{7.70} &
  \multicolumn{1}{r|}{6.07} &
  {[}-1.99, 1.43{]} \\ \hline
\textit{slower} &
  \multicolumn{1}{r|}{\textbf{4.35}} &
  \multicolumn{1}{r|}{7.75} &
  \multicolumn{1}{r|}{7.31} &
  {[}-2.33, 1.59{]} \\ \hline
\textit{veryslow} &
  \multicolumn{1}{r|}{\textbf{5.14}} &
  \multicolumn{1}{r|}{5.68} &
  \multicolumn{1}{r|}{7.90} &
  {[}-2.13, 2.13{]} \\ \hline 
\textbf{average} &
  \multicolumn{1}{r|}{\textbf{4.89}} &
  \multicolumn{1}{r|}{6.37} &
  \multicolumn{1}{r|}{5.37} &
  \multicolumn{1}{c|}{-} \\ \hline 
\end{tabular}
\vspace{-0.25cm}
\end{table}
Furthermore, to assess the performance of the proposed energy estimator models, we used a confidence interval analysis on the estimation errors, i.e., percentage errors. To this end, we obtained the confidence interval $CI_{X}$ for a 95\% confidence level with ${m}_{X}$ being mean and ${\sigma}_{X}$ being standard deviation of the percentage errors of all the B bit streams associated with preset $X$, $r_{X}=[r_{X,1},...,r_{X,B}]$. Then, for each preset $X$, and $B$ bit streams, the confidence interval is defined as follows:
\begin{equation}
    CI_{X}={m}_{X} \pm z \cdot ({\sigma}_{X}/\sqrt{B}),
\end{equation}
where z is the z-score corresponding to the chosen confidence level, which is  1.96 for 95\% confidence level. In Table \ref{tab:posterior} and Table \ref{tab:prior}, we report the MAPE for each preset, the average MAPE over all the presets, and the confidence intervals of the percentage errors for all the presets of the proposed models. In subsequent subsections, we evaluate the posterior and prior estimation models, followed by a discourse on their practical applications. 
\vspace{-0.25cm}
\subsection{Posterior Estimation Models}
\vspace{-0.25cm}
Firstly, we evaluate the proposed PE-based posterior estimation model \eqref{complexity} in comparison with the models from the literature \cite{Ramasubbu22} and \cite{Ramasubbu22_2}. The PE-based posterior estimation model achieves a MAPE of less than 10\% for all the presets. In addition, the confidence intervals of the percentage errors include zero, suggesting that the proposed model's estimations are unbiased, i.e., the model does not systematically underestimate or overestimate the encoding energy values. The MAPE values in Table \ref{tab:posterior} show that the PE-based proposed posterior estimation model outperforms the BSF model but outperforms the time-based posterior estimation model only in faster presets. PE-based and time-based posterior estimation models perform best. However, the time-based posterior estimation model is generally better because of the lower complexity, as the runtime of the PE-based posterior estimation model is more than that of the time-based posterior estimation model. Nevertheless, the posterior PE-based estimation model can be used to investigate the energy consumption of various encoding sub-processes in detail, which, in turn, aids in obtaining function-specific energy demand, as measuring such function-specific energy demand is practically impossible.

\begin{figure}[]
\centering
 \input{Randplot.tex} \includegraphics[width=0.45\textwidth]{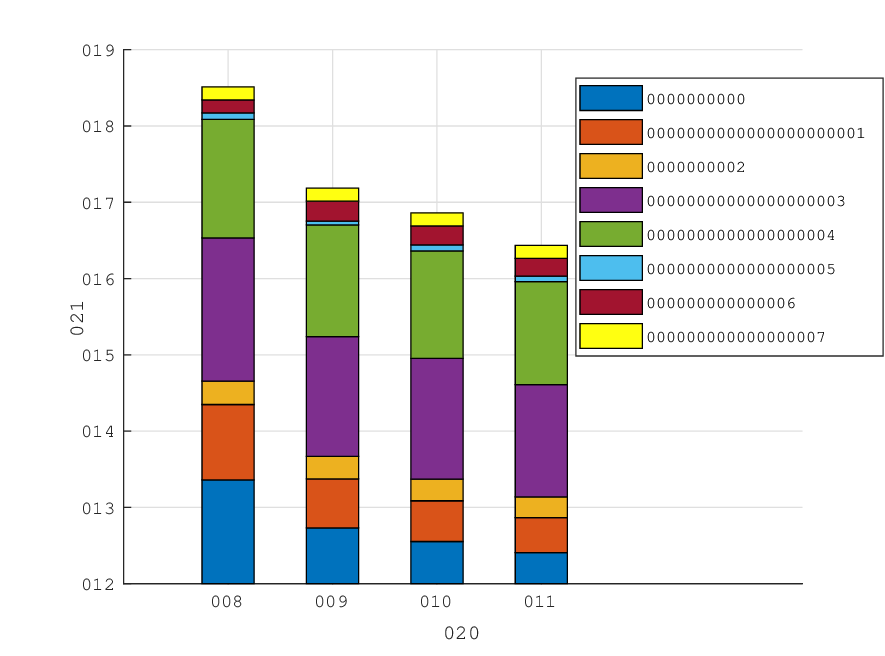}
  \caption{The estimated energy $\hat E_\mathrm{enc}$ using \eqref{complexity}, with the energy distribution for various sub-processes, for a single frame class B sequence, "BasketballDrive," using x265 encoder, at medium preset and CRF values of 18, 23, 28, and 33.}
 \label{fig:distribution}
 \vspace{-0.45cm}
\end{figure}
The proposed PE-based posterior estimation model has a significant application in analyzing the energy demand distribution among different encoding sub-processes. By employing the proposed PE-based posterior estimation model \eqref{complexity}, we can obtain the PE-specific energies $e_{\text{\textit{i}}}$ of the encoding process from its corresponding profile events and encoding energy. Furthermore, Valgrind, in its profiling capacity, not only furnishes global profiling outcomes but also executes profiling for individual encoding sub-processes. Consequently, by combining these sub-process-specific PEs with PE-specific energies $e_{\text{\textit{i}}}$, we can achieve a comprehensive understanding of the energy requirements across different encoding sub-processes. Figure \ref{fig:distribution} illustrates the energy distribution of encoding a single frame from a class B sequence, "BasketballDrive," using the x265 encoder at medium preset and CRF values of 18, 23, 28, and 33.

Further analysis of the energy distribution of intra-coding, illustrated in Figure \ref{fig:distribution}, reveals that the intra-mode search constitutes the most energy-intensive aspect, contributing between 28.5\% and 33.19\% of the total energy demand across the CRF range from 18 to 33. Following closely, intra-prediction accounts for 23.86\% to 30.42\% of the total energy demand. Subsequently, entropy coding contributes 20.6\% of the total energy demand at CRF 18. In addition, its contribution decreases with increased CRF values, as the number of coefficients to be entropy-coded decreases with higher CRF values. Then, the quantization and transform coding consume 15\% and 10\% of total encoding energy for CRF 18 and 33, respectively. Energy consumption by in-loop filters ranges from 4.62\% to 6.11\% of the total energy demand across the CRF range of 18 to 33. Additionally, global initialization, as well as CTU and CU level preprocessing, on average, account for approximately 4\%, 3\%, and 1.3\% of the total energy consumption, respectively.

In summary, using \eqref{complexity} to obtain energy distribution of the encoding process offers a versatile set of applications. It can help identify major and significant energy contributors in the encoding process, compare the energy distribution of the encoding process of different encoding presets, and compare the difference in energy distribution for different sequences.

\begin{table}[]
\small
\centering
\caption{MAPE for all the prior estimation models and the confidence interval of percentage errors of the proposed model. The minimum MAPE across all models are in bold.}
\label{tab:prior}
\begin{tabular}{|l|rrrl|}
\hline
\multirow{3}{*}{\textbf{\begin{tabular}[c]{@{}l@{}}Encoder \\ Preset\end{tabular}}} &
  \multicolumn{4}{c|}{\textbf{Prior Estimation Models}} \\ \cline{2-5} 
 &
  \multicolumn{1}{c|}{\textbf{\begin{tabular}[c]{@{}c@{}}QP\\ \cite{Rodriguez15}\end{tabular}}} &
  \multicolumn{1}{c|}{\textbf{\begin{tabular}[c]{@{}c@{}}UF Time\\ \cite{Ramasubbu22}\end{tabular}}} &
  \multicolumn{2}{c|}{\textbf{\begin{tabular}[c]{@{}c@{}}UF PE\\ \eqref{complexityUF}\end{tabular}}} \\ \cline{2-5} 
 &
  \multicolumn{3}{c|}{\textbf{MAPE, }$\boldmath{{\epsilon}_{X}}$ \textbf{(\%)}} &
  \multicolumn{1}{c|}{\boldmath{$CI_{X}$}\textbf{(\%)}} \\ \hline
\textit{ultrafast} &
  \multicolumn{1}{r|}{17.98} &
  \multicolumn{1}{r|}{-} &
  \multicolumn{1}{r|}{-} &
  \multicolumn{1}{c|}{-} \\ \hline
\textit{superfast} &
  \multicolumn{1}{r|}{22.30} &
  \multicolumn{1}{r|}{5.68} &
  \multicolumn{1}{r|}{\textbf{3.89}} &
  {[}-1.3, 0.97{]} \\ \hline
\textit{veryfast} &
  \multicolumn{1}{r|}{21.00} &
  \multicolumn{1}{r|}{6.90} &
  \multicolumn{1}{r|}{\textbf{4.70}} &
  {[}-1.51, 1.06{]} \\ \hline
\textit{faster} &
  \multicolumn{1}{r|}{21.54} &
  \multicolumn{1}{r|}{6.82} &
  \multicolumn{1}{r|}{\textbf{4.64}} &
  {[}-1.5, 1.05{]} \\ \hline
\textit{fast} &
  \multicolumn{1}{r|}{19.34} &
  \multicolumn{1}{r|}{5.83} &
  \multicolumn{1}{r|}{\textbf{4.61}} &
  {[}-1.49, 1.08{]} \\ \hline
\textit{medium} &
  \multicolumn{1}{r|}{23.67} &
  \multicolumn{1}{r|}{10.14} &
  \multicolumn{1}{r|}{\textbf{5.34}} &
  {[}-1.71, 1.12{]} \\ \hline
\textit{slow} &
  \multicolumn{1}{r|}{24.02} &
  \multicolumn{1}{r|}{13.35} &
  \multicolumn{1}{r|}{\textbf{6.44}} &
  {[}-2.07, 1.28{]} \\ \hline
\textit{slower} &
  \multicolumn{1}{r|}{25.81} &
  \multicolumn{1}{r|}{24.96} &
  \multicolumn{1}{r|}{\textbf{6.94}} &
  {[}-2.62, 0.96{]} \\ \hline
\textit{veryslow} &
  \multicolumn{1}{r|}{29.17} &
  \multicolumn{1}{r|}{28.82} &
  \multicolumn{1}{r|}{\textbf{7.33}} &
  {[}-2.71, 1.02{]} \\ \hline
\textbf{average} &
  \multicolumn{1}{r|}{22.65} &
  \multicolumn{1}{r|}{12.05} &
  \multicolumn{1}{r|}{\textbf{5.36}} &
  \multicolumn{1}{c|}{-} \\ \hline
\end{tabular}
\end{table}
\vspace{-0.25cm}
\subsection{Prior Estimation Models}
Secondly, we evaluate the proposed UF PE-based prior estimation model, which allows estimation of the encoding energy without performing actual encoding for presets slower than \textit{ultrafast}. The proposed UF PE-based estimation model \eqref{complexityUF} yields a MAPE of 5.36\% and, when the preset is known, outperforms the QP-based model and UF time-based model with a MAPE of 22.65\% and 12.05\% from the literature. Additionally, the confidence intervals of the percentage errors of the proposed model, as shown in Table \ref{tab:prior}, indicate that the proposed model's estimations are unbiased, similar to the observation from Table \ref{tab:posterior}. Regarding computational costs, the proposed UF PE-based prior estimation model performs better than the PE-based posterior estimation model for practical estimation. Unlike the proposed posterior estimation model, this prior estimator requires the PEs of the lightweight encoding configuration UF preset and can be obtained with less complex preprocessing overhead compared to the other presets. Therefore, such a model could be used for practical energy estimations. While the PE-based prior estimation model yields accurate estimations, the time-based prior estimation model offers lower computational complexity. Specifically, the runtime of the UF PE-based prior estimation model is five times longer than that of the UF time-based counterpart. Nonetheless, for precise estimation of encoding energy consumption, the UF PE-based prior estimation model proves most beneficial, particularly for slower presets such as \textit{medium}, \textit{slow}, \textit{slower}, and \textit{veryslow}.


\begin{figure}[]
\centering
%
\providecommand\matlabfragNegTickNoWidth{\makebox[0pt][r]{\ensuremath{-}}}%
%
%
\providecommand\matlabtextA{\color[rgb]{0.000,0.000,0.000}\fontsize{6.35}{6.35}\selectfont\strut}%
\psfrag{000000000000000000000000000}[cl][cl]{\matlabtextA Measured Encoding Energy}%
\psfrag{00000000000000000000000000000000000001}[cl][cl]{\matlabtextA Estimated Encoding Energy (UF Time)}%
\psfrag{000000000000000000000000000000000002}[cl][cl]{\matlabtextA Estimated Encoding Energy (UF PE)}%
\providecommand\matlabtextB{\color[rgb]{0.150,0.150,0.150}\fontsize{9.00}{9.00}\selectfont\strut}%
\psfrag{018}[tc][tc]{\matlabtextB x265 presets}%
\psfrag{019}[bc][bc]{\matlabtextB \begin{tabular}{@{}c@{}}~Energy~per~100k~pixels~~~~~~\\~~~~in~Joules~~~~~~~~~~~~~~~~\end{tabular}}%
%
%
%
\providecommand\matlabtextC{\color[rgb]{0.150,0.150,0.150}\fontsize{7.00}{7.00}\selectfont\strut}%
\psfrag{003}[ct][ct]{\matlabtextC $1$}%
\psfrag{004}[ct][ct]{\matlabtextC $2$}%
\psfrag{005}[ct][ct]{\matlabtextC $3$}%
\psfrag{006}[ct][ct]{\matlabtextC $4$}%
\psfrag{007}[ct][ct]{\matlabtextC $5$}%
\psfrag{008}[ct][ct]{\matlabtextC $6$}%
\psfrag{009}[ct][ct]{\matlabtextC $7$}%
\psfrag{010}[ct][ct]{\matlabtextC $8$}%
\psfrag{011}[ct][ct]{\matlabtextC $9$}%
%
%
%
\psfrag{012}[rc][rc]{\matlabtextC $0$}%
\psfrag{013}[rc][rc]{\matlabtextC $0.5$}%
\psfrag{014}[rc][rc]{\matlabtextC $1$}%
\psfrag{015}[rc][rc]{\matlabtextC $1.5$}%
\psfrag{016}[rc][rc]{\matlabtextC $2$}%
\psfrag{017}[rc][rc]{\matlabtextC $2.5$}%
%
 \includegraphics[width=0.45\textwidth]{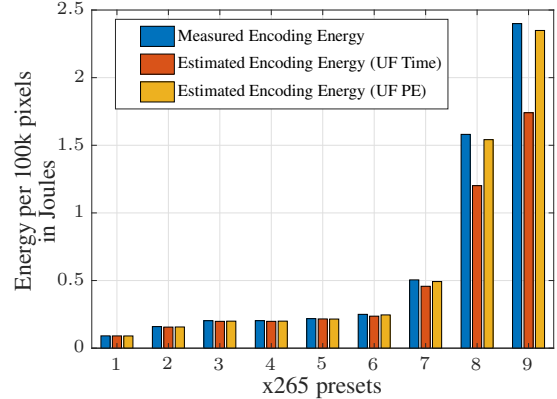}
   \vspace{-0.25cm}
  \caption{The measured encoding energy $E_\mathrm{enc}$, and estimated encoding energy $\hat E_\mathrm{enc}$ using UF Time \cite{Ramasubbu22}, and estimated encoding energy $\hat E_\mathrm{enc}$ using UF PEs \eqref{complexityUF}, for 100000 pixels, averaged over all the sequences, for the CRF value 23, where ${1,2,3,4,5,6,7,8,9}$ corresponds to the x265 presets.}
  \vspace{-0.45cm}
 \label{fig:estimation}
\end{figure}
The proposed PE-based prior estimation model has a significant application in accurately estimating the energy demand of the encoding process compared to existing literature models. In Figure \ref{fig:estimation}, the encoding energy is illustrated: blue bars represent measured values, orange bars represent estimates using the UF (\textit{Ultrafast}) Time-based prior estimation model \cite{Ramasubbu22}, and yellow bars represent estimates using the UF PE-based prior estimation model. These measured energy and estimates are provided for various x265 presets, including \textit{ultrafast}, \textit{superfast}, \textit{veryfast}, \textit{faster}, \textit{fast}, \textit{medium}, \textit{slow}, \textit{slower}, and \textit{veryslow}, numbered 1 through 9 respectively. The analysis of the measured encoding energy reveals relative differences in energy demand between presets: \textit{superfast} consumes 75\% more energy than \textit{ultrafast}; \textit{veryfast} consumes 27\% more energy than \textit{superfast}; \textit{faster} requires 0.1\% more energy than \textit{veryfast}; \textit{fast} demands 7\% more energy than \textit{faster}; \textit{medium} requires 15\% more energy than \textit{fast}; \textit{slow} consumes 102\% more energy than \textit{medium}; \textit{slower} consumes 213\% more energy than \textit{slow}; and \textit{veryslow} requires 52\% more energy than \textit{slower}. The data depicted in Figure \ref{fig:estimation} illustrates that both the UF-time based prior estimation model \cite{Ramasubbu22} and the UF PE-based prior estimation model offer precise estimates for presets ranging from \textit{ultrafast} to \textit{fast}. However, when it comes to slower presets such as \textit{medium}, \textit{slow}, \textit{slower}, and \textit{veryslow}, the UF PE-based prior estimation model outperforms the UF-time based model, providing the more accurate estimations. Consequently, for encoding energy estimates, the UF Time-based estimation model suffices for faster presets such as \textit{ultrafast}, \textit{superfast}, \textit{veryfast}, \textit{faster}, and \textit{fast}, whereas the proposed prior estimation model proves beneficial for slower presets.
\vspace{-0.25cm}
\section{Conclusion}
\label{sec:concl}
\vspace{-0.25cm}
Energy measurements are pivotal in developing energy-efficient video coding algorithms. Nonetheless, such measurements are complex and costly. Thus, we need valid and simple energy estimation models. This work demonstrates that, for the HEVC software encoding process, a lightweight PE-based posterior estimation model provides an accurate prior estimation of encoding energy, exhibiting a mean absolute percentage error of 5.36\%. Moreover, this work also presents the PE-based posterior modeling approach, which yields the average sub-process-specific energies, aiding in analyzing the energy demand of encoding sub-processes. In addition, we presented an examplary energy consumption analysis of the HEVC encoding process on a functional level using the PE-based posterior model. In the future, we plan to extend the proposed models to other encoders and extend the energy consumption analysis of the encoding process on a functional level, such as motion estimation and compensation.
\vspace{-0.25cm}

\bibliographystyle{IEEEbib}
\begin{small}
\vspace{-0.25cm}
    \bibliography{literature}
\end{small}

\end{document}